\documentclass[showpacs,twocolumn,floats,prb]{revtex4}
\usepackage{amssymb}

\usepackage{graphicx}

\begin{document}

\title{Comment on "Composite excitation of Josephson phase and spin waves in
ferromagnetic Josephson junctions" (S.Hikino, M.Mori, S.Takahashi, and
S.Maekawa, arXiv:cond-mat 1009.3551)}
\author{A.F. Volkov${}^{1,2}$}
\affiliation{$^{1}$Theoretische Physik III,\\
Ruhr-Universit\"{a}t Bochum, D-44780 Bochum, Germany\\
$^{2}$Institute for Radioengineering and Electronics of Russian Academy of\\
Sciences,11-7 Mokhovaya str., Moscow 125009, Russia}

\begin{abstract}
We clarify the applicability of the quasistatic approximation used in Ref.
[\onlinecite{VE}], where coupled spin and Josephson plasma waves have been predicted
to exist in SIFS Josephson junctions. We show, contrary to the claim of the
authors of Ref. [\onlinecite{Maekawa}], that this approximation is very accurate in realistic
systems studied experimentally.
\end{abstract}

\maketitle

\bigskip

The dynamics of SIFS (or SFIFS) Josephson junctions with ferromagnetic layers has been studied 
theoretically in Ref. [\onlinecite{VE}]. Here S, I and F stand
for a superconducting, thin insulating or ferromagnetic layer, respectively. In
particular, it has been shown that weakly damped magneto-plasma oscillations are
possible in such a system. That is, oscillations of the magnetic moment \ $M$
in the F layer and Josephson ''plasma'' waves turn out to be coupled. The
coupled modes (spin and Josephson plasma waves) may result in
the peaks on the I-V characteristics of the junction in addition
to the Fiske steps.

The same problem has been studied in a recent paper [\onlinecite{Maekawa}]. The authors
claim that the electromagnetic (EM) fields in the F layer which excite spin
waves in F have been neglected in Ref. [\onlinecite{VE}].

\bigskip In this Comment, we would like to clarify that:

A) contrary to the statement of Ref. [\onlinecite{Maekawa}], the EM fields 
 in the F film are taken into account in Ref. [\onlinecite{VE}]. Indeed,
 there could be no coupling between magnetic and plasma modes otherwise.

B) these fields are considered in the quasistatic approximation, which describes the dynamics 
of realistic junctions rather accurately, while the effects of ac electric fields $E$, accounted for
in Ref. [\onlinecite{Maekawa}], is negligible. The only important ac field is the magnetic induction
$B$ which is described by the London equation. The skin effect due to quasiparticle
current driven by $E$ can be neglected.

\bigskip Estimations justifying this approximation are not presented in Ref. [\onlinecite{VE}]
for lack of space. Here we give these simple estimations and present a
physical explanation of the coupling between the spin and Josephson plasma
waves.

The equation for the phase difference $\varphi $ (Eq.(4) in Ref. [\onlinecite{VE}]) is
obtained from the Maxwell equation

\begin{equation}
(\mathbf{\nabla }\times \mathbf{B})_{z}=\frac{4\pi }{c}j_{z}  \label{1}
\end{equation}%
written in the superconducting regions S (where the magnetic induction $%
\mathbf{B}$ coincides with the magnetic field $\mathbf{H}$) and from the
usual expression for the current through the Josephson junction. The
displacement current $j_{dis}=(\epsilon /c)\partial E/\partial t$ is dropped
because in metals it is negligible in comparison with the quasiparticle
current ($\omega \ll \sigma _{Q}$), where at $T\precsim \Delta $ the
quasiparticle conductance $\sigma _{Q}\approx \sigma _{Dr}\exp (-\Delta /T)$
with $\sigma _{Dr}=(e^{2}n\tau /m)\approx 10^{17}s^{-1}$ for the mean free
path $l=v\tau \approx 10^{-6}cm.$

\bigskip In the quasistatic approximation the expression for $\mathbf{B(}z,t%
\mathbf{)}$ is given by Eq.(3) in Ref. [\onlinecite{VE}]

\begin{equation}
\mathbf{B}_{\perp}(z,t)=\{\frac{\Phi _{0}}{4\pi \lambda _{L}}\mathbf{n}_{z}%
\mathbf{\times \nabla }_{\perp }\varphi -\frac{2\pi \tilde{d}_{F}}{\lambda
_{L}}\mathbf{M}_{\perp }\}\exp (-\frac{(z-d_{F})}{\lambda _{L}}).  \label{2}
\end{equation}

It relates the magnetic field in the superconductors and the phase
difference $\varphi (t).$ The second term in the curly brackets appears due to
the magnetic moment in the F layer(s). Integrating this expression
over the square of the superconductors (in the $(x,z)$-plane perpendicular to the magnetic field)
and adding the magnetic moment of
the F layer $4\pi \tilde{d}_{F}L_{x}M,$,  we obtain the usual law of the
quantization of the magnetic moment in Josephson junctions: $\Phi \equiv
\Phi _{S}+\Phi _{F}=\Phi _{0}n$, where $L_{x}$ is the length of the superconductors in the $x$-direction,
 $\Phi _{0}$ is the magnetic flux
quantum and $n$ is an integer. Due to the second term in Eq.(2) the
Josephson mode is coupled to the spin waves.

Eq.(\ref{2}) for $\mathbf{B}_{\perp }$ is obtained by using expression for
the current $\mathbf{j}_{\perp }$ (Eq.(1) in Ref. [\onlinecite{VE}]) which is written
in the dirty limit ($\omega \tau \ll 1,kl\ll 1,$ where $\omega ,k$ are
characteristic frequency and wave vector, respectively, $l=v\tau $ is the
mean free path). The quasiparticle current $\mathbf{j}_{Q\perp }=\sigma
_{Q}(\omega )\mathbf{E}_{\perp }$ and therefore the transverse electric
field $\mathbf{E}_{\perp }$ ($\mathbf{\nabla }\times \mathbf{E}_{\perp }\neq
0$) is neglected. This approximation is valid if the skin depth $\delta _{sk}
$ is much larger than the London penetration length $\lambda _{L}.$ If the
current $\mathbf{j}_{Q\perp }$ is taken into account, then $\lambda _{L}$ in
Eq.(3) of Ref. [\onlinecite{VE}] should be replaced by $\lambda _{\omega }=1/\sqrt{%
\lambda _{L}^{-2}+4\pi i\omega \sigma _{Q}/c^{2}}$. The second term is small
in comparison with the first one if the frequency $\omega =2\pi \nu $ is not
very high

\begin{equation}
\nu \ll \frac{1}{8\pi ^{2}\sigma _{Q}}(\frac{c}{\lambda _{L}})^{2}\approx
5\cdot 10^{12}\exp (\frac{\Delta }{T})Hz  \label{C2}
\end{equation}%
where we take $\lambda _{L}\approx 5\cdot 10^{-6}cm$. For the realistic SIFS
junctions, where the frequency $\nu $ typically is less than one hundred gigahertz [\onlinecite{Weides}],
this condition is easily fulfilled.

The currents induced in the F layer also change the magnetic induction $%
\mathbf{B}$. However this change, $\delta \mathbf{B}_{F}$, is even smaller
than the change $\delta \mathbf{B}_{SQ}$ caused by the quasiparticle current
in S. Indeed, the change $\delta \mathbf{B}_{F}$ is determined by the total
Meissner current in the F layer $j_{FMeis}d_{F}$ which is much smaller than $%
j_{SMeis}\lambda _{L}$ because $d_{F}\ll \lambda _{L}$ (by assumption) and $%
j_{FMeis}\ll j_{SMeis}$ since the condensate density in F is significantly lower than
the density of Cooper pairs in S.

The change $\delta \mathbf{B}_{F}$ due to skin effect can be neglected if
the frequency satisfies the condition

\begin{equation}
\nu \ll \frac{1}{8\pi ^{2}\sigma _{F}}(\frac{\lambda _{L}}{d_{F}})(\frac{c}{%
\lambda _{L}})^{2}  \label{C3}
\end{equation}%
This condition is fulfilled even easier than the condition (2).

Therefore, the currents and ac electric fields in F accounted for in 
Ref. [\onlinecite{Maekawa}] can be neglected. The only EM field
in the F layer, which is essential, is the induction $\mathbf{B}$ determined by Eq.(3)
of Ref. [\onlinecite{VE}]. The quasistatic approximation used in deriving this
equation is fulfilled for realistic systems (see Ref. [\onlinecite{Weides}]) with a
great accuracy.

We thank SFB 491 for financial support.

\end{document}